\documentstyle[11pt]{article}
\begin{document}

\title{ Single chargino production via gluon-gluon fusion
        in a supersymmetric theory with an explicit R-parity violation
\footnote{Supported by National Natural Science Foundation of
          China.}}
\vspace{3mm}

\author{{ Wan Lang-Hui$^{b}$, Ma Wen-Gan$^{a,b}$, Yin Xi$^{b}$,
          Jiang Yi$^{b}$ and Han Liang$^{b}$  }\\
{\small $^{a}$ CCAST (World Laboratory), P.O.Box 8730, Beijing 100080,P.R.China} \\
{\small $^{b}$ Department of Modern Physics, University of Science and Technology}\\
{\small of China (USTC), Hefei, Anhui 230027, P.R.China} }

\date{}
\maketitle

\vskip 12mm

\begin{abstract}

We studied the production of single chargino $\tilde{\chi}_1^{\pm}$ accompanied
by $\mu^{\mp}$ lepton via gluon-gluon fusion at the LHC. The numerical
analysis of their production rates is carried out in the mSUGRA scenario
with some typical parameter sets. The results show that the cross sections
of the $\tilde{\chi}_1^{\pm}\mu^{\mp}$ productions via gluon-gluon collision
are in the order of $1 \sim 10^{2}$ femto barn quantitatively at the CERN LHC,
and can be competitive with production mechanism via quark-antiquark
annihilation process.

\end{abstract}

\vskip 5cm

{\large\bf PACS: 13.88.+e, 13.65.+i, 14.80.Dq, 14.65.-q, 14.80.Gt}

\vfill \eject

\baselineskip=0.36in

\renewcommand{\theequation}{\arabic{section}.\arabic{equation}}
\renewcommand{\thesection}{\Roman{section}}

\makeatletter      
\@addtoreset{equation}{section}
\makeatother       

\section{Introduction}
\par
Over the past years intensive investigations into the new physics beyond
the standard model (SM) has been undertaken\cite{r1}. As the simplest extension
of the SM, the supersymmetric model (SUSY) is the most attractive one.
In general extended models of the SM, electroweak gauge invariance forbids
terms in the SM Lagrangian that change either baryon number or lepton number,
such terms are allowed in the most general supersymmetric (SUSY) extension of
the SM, but they may lead to an unacceptable short proton lifetime, One way
to evade the proton-decay problem is to impose a discrete symmetry
conservation called R-parity ($R_{p}$) conservation.
In this case, all supersymmetric partner particles must
be pair-produced, thus the lightest of superparticles must be stable.
\par
The R-parity violation ($\rlap/R_{p}$) implies either lepton number or baryon
number being broken, and it will change the feature of the SUSY models a lot.
Due to the lack of experimental tests for $R_{p}$ conservation, the $R_{p}$
violation case is also equally well motivated in the supersymmetric extension
of the SM. SUSY models with $\rlap/R_{p}$ can provide many interesting
phenomena. Recently there are some investigations on the signal on $R_{p}$
violation \cite{s2}\cite{s3}\cite{s4}, because of experimentally observed
discrepancies.
\par
In the last few years, many efforts were made to find $\rlap/R_{p}$
interactions in experiments. Unfortunately, up to now we have only some
upper limits on $\rlap/R_{p}$ parameters, such as B-violating
$\rlap/R_{p}$ parameters ($\lambda^{''}$) and L-violating $\rlap/R_{p}$
parameters($\lambda$ and $\lambda^{'}$)\cite{s4}\cite{s5}\cite{s6}
(The parameters will be defined clearly in the following sector). Therefore,
trying to find the signal of $R_{p}$ violation or getting more stringent
constraints on the parameters in future experiments, is a promising
task. The popular way to find a $R_{p}$ violation signal
is to detect the decay of the lightest supersymmetric particle(LSP)\cite{s4}
\cite{s6}\cite{s7}, but it is difficult experimentally especially
at hadron colliders. The best signal for $\rlap/R_{p}$ at the CERN Large
Hadron Collider (LHC) is the resonant sneutrino production through a
$\lambda^{'}$ or a $\lambda^{''}$ coupling constant, respectively\cite{Moreau}.
Because the c.m.s energy continuous distribution of the colliding partons
inside protons at the hadron colliders, a intermediate resonance can be
probed over a rather wide mass range. Then a single chargino can be produced
by the sneutrino decays, which can be measured through the detection of its
three-leptons signature.
\par
There are two mechanisms at parton level to produce $\tilde{\chi}_{1}^{\pm}
\mu^{\mp}$ in an explicit R-parity violating SUSY theory at pp colliders.
One is via quark and antiquark
($q\bar{q}(q=u,d)$) annihilation which is allowed at tree-level. The single
lightest chargino production at the LHC as induced by the resonant sneutrino
production $pp \rightarrow q\bar{q}\rightarrow \tilde{\nu}_{\mu} \rightarrow
\tilde{\chi}_{1}^{\pm} \mu^{\mp}$ was studied by G. Moreau et.al.\cite{Moreau}.
Another mechanism is via gluon-gluon fusion. The single lightest chargino
production process via gluon-gluon fusion, which will take place at the
lowest order by one-loop diagrams, could be also significant due to large
gluon luminosity in distribution function of proton.
\par
In this paper, we investigate the resonant sneutrino particle production
via gluon-gluon fusion at the LHC operating at the energy of $14~TeV$.
We arrange this paper as follows. In Sec.II we present the analytical
calculations of both subprocess and parent process. In Sec.III we give
some numerical presentations in the MSSM and the minimal
supergravity (mSUGRA) scenario \cite{msugra}, and discuss these numerical
results. The conclusions are contained in Sec.IV. Finally some notations used
in this paper, the explicit expressions of the form factors induced by
the loop diagrams are collected in Appendix.

\par
\section{The Calculation of $pp \rightarrow
    gg \rightarrow  \tilde{\chi}_1^{\pm}\mu^{\mp}+X $}
\par
The R-parity of a particle is defined as $R_{p}=(-1)^{2S+3B+L}$ \cite{r2},
where $S$ is the spin quantum number of the particle, $L$ the lepton number
and $B$ the baryon number. The minimal supersymmetric model (MSSM) does not
contain the most general superpotential respecting to the gauge symmetries of
the SM, which includes bilinear and trilinear terms, which can be expressed as
\begin{equation}
\label{sup}
    W_{\rlap/R_{p}} =\frac{1}{2}
\lambda_{[ij]k} L_{i}.L_{j}\bar{E}_{k}+\lambda^{'}_{ijk}
L_{i}.Q_{j}\bar{D_{k}}+\frac{1}{2}\lambda^{''}_{i[jk]}
\bar{U}_{i}\bar{D}_{j}\bar{D}_{k}+\epsilon _{i} L_{i} H_{u}.
\end{equation}
where $L_i$, $Q_i$ are the SU(2) doublet lepton and quark
fields, $E_i$,$U_i$,$D_i$ are the singlet superfields. The $UDD$
couplings violate baryon number while the other three sets violate lepton
number. We shall explicitly forbid the $UDD$ interactions as an economical
way to avoid unacceptable rapid proton decay\cite{Ross}, and the term of
$LLE$ have no contribution to our process $pp \rightarrow gg \rightarrow
\tilde{\chi}_1^{+} \mu^{-}+X$, we shall not discuss them too.
In this work we ignored the bilinear terms that mix lepton and Higgs superfields
\cite{s3} for simplicity, because its effects are small in our process.

\par
Expanding the superfield components in Eq.(\ref{sup}) we obtain the interaction
Lagrangian that contains quarks and leptons:
\begin{equation}
\label{lag}
{\cal L}_{LQD}=\lambda_{ijk}^{'}\{\tilde{\nu}_{iL}\bar{d}_{kR}d_{jL}-
      \tilde{e}_{iL}\bar{d}_{kR}u_{jL}+\tilde{d}_{jL}
      \bar{d}_{kR}\nu_{iL}- \tilde{u}_{jL}\bar{d}_{kR}e_{iL}+
      \tilde{d}_{kR}^c\nu_{iL}d_{jL}- \tilde{d}_{kR}^ce_{iL}u_{jL}\} + h.c.
\end{equation}
\par
The subprocess $gg \rightarrow \tilde{\chi}_1^{\pm}\mu^{\mp}$ can only be
produced through one-loop diagram in the lowest order. In this case it is
not necessary to consider the renormalization. The generic Feynman diagrams
contributing to the subprocess in the MSSM without R-parity at one-loop level
are depicted in Fig.1, where the exchange of incoming gluons in
Fig.1$(a.1\sim3)$ and Fig.1$(c.1,2)$ are not shown. We divide all the one-loop
diagrams in Fig.1 into three groups: (1) box diagrams shown in
Fig.1(a), (2) quartic interaction diagrams in Fig.1(b),
(3) triangle diagrams shown in Fig.1(c). In this work, we perform the
calculation in the 't Hooft-Feynman gauge. The relevant Feynman rules
without $\rlap/{R}_{p}$ interactions can be found in references\cite{r1}
\cite{Gunion}\cite{Mwg}. The related Feynman rules with $\rlap/{R}_{p}$
interactions can be read out from Eq.(\ref{lag}).
\par
If we ignore the CP violation, the cross section of $pp \rightarrow gg
\rightarrow \tilde{\chi}_1^+ \mu^- +X$ coincides with the process $pp
\rightarrow gg \rightarrow \tilde{\chi}_1^- \mu^+$ because of charge
conjugation invariance and we shall specify on the calculation of the
$\tilde{\chi}_1^+ \mu^-$ production for simplicity in the following.
We denote the reaction of $\tilde{\chi}_1^+$ and $\mu^-$ production via
gluon-gluon fusion as:
\begin{equation}
g(p_1, \alpha, \mu) g(p_2, \beta, \nu) \longrightarrow
\tilde{\chi}_1^+ (k_1) \mu^- (k_2).
\end{equation}
where $p_1$ and $p_2$ denote the four momenta of the incoming gluons,
$k_1$, $k_2$ denote the four momenta of the outgoing chargino and $\mu^-$
lepton respectively, and $\alpha$, $\beta$ are color indices of the colliding
gluons.
\par
The corresponding matrix element of Feynman diagrams in Fig.1, can be written
as
\begin{eqnarray*}
{\cal M}&=& {\cal M}^{b}+{\cal M}^{q}+{\cal M}^{tr}=\epsilon^{\mu}(p1)\epsilon^{\nu}(p2)
\bar{u}(k_1)\{
 f_1g_{\mu\nu}
 + f_2g_{\mu\nu} \gamma_5
 + f_3k_{1\mu}k_{1\nu}
 +  f_4k_{1\mu}k_{1\nu} \gamma_5 \\
 &+& f_5k_{1\nu} \gamma_{\mu}
 + f_6k_{1\mu} \gamma_{\nu}
 + f_7g_{\mu\nu} \rlap/{p}_1
 + f_8k_{1\mu}k_{1\nu} \rlap/{p}_1
 + f_9g_{\mu\nu} \rlap/{p}_2
 + f_{10}k_{1\mu}k_{1\nu} \rlap/{p}_2
 + f_{11}k_{1\nu} \gamma_5 \gamma_{\mu}   \\
 &+& f_{12}k_{1\mu} \gamma_5\gamma_{\nu}
 + f_{13}g_{\mu\nu} \gamma_5\rlap/{p}_1
 + f_{14}k_{1\mu}k_{1\nu} \gamma_5\rlap/{p}_1
 + f_{15}g_{\mu\nu} \gamma_5\rlap/{p}_2
 + f_{16}k_{1\mu}k_{1\nu} \gamma_5\rlap/{p}_2 \\
 &+& f_{17}k_{1\nu} \gamma_{\mu} \rlap/{p}_1
 + f_{18}k_{1\nu} \gamma_5\gamma_{\mu}\rlap/{p}_1
 + f_{19} \gamma_{\mu}\gamma_{\nu}\rlap/{p}_1
 + f_{20}k_{1\nu} \gamma_{\mu}\rlap/{p}_1\rlap/{p}_2
 + f_{21} \gamma_5\gamma_{\mu}\gamma_{\nu}\rlap/{p}_1\\
 &+& f_{22}k_{1\nu} \gamma_5 \gamma_{\mu}\rlap/{p}_1\rlap/{p}_2
 + f_{23}k_{1\mu} \gamma_{\nu}\rlap/{p}_2
 + f_{24}k_{1\mu} \gamma_5 \gamma_{\nu}\rlap/{p}_2
 + f_{25} \gamma_{\mu} \gamma_{\nu}\rlap/{p}_2
 + f_{26}k_{1\mu} \gamma_{\nu} \rlap/{p}_1\rlap/{p}_2  \\
 &+& f_{27} \gamma_5 \gamma_{\mu}\gamma_{\nu}\rlap/{p}_2
 + f_{28}k_{1\mu} \gamma_5\gamma_{\nu}\rlap/{p}_1\rlap/{p}_2
 + f_{29}\epsilon_{\mu \nu \alpha \beta} p_1^\alpha p_2^\beta
 + f_{30}\epsilon_{\mu \nu \alpha \beta} p_1^\alpha p_2^\beta \gamma_5
 + f_{31} \gamma_{\mu}\gamma_{\nu} \\
 &+& f_{32} \gamma_5 \gamma_{\mu}\gamma_{\nu}
 + f_{33} \gamma_{\mu} \gamma_{\nu}\rlap/{p}_1\rlap/{p}_2
 + f_{34} \gamma_5\gamma_{\mu}\gamma_{\nu}\rlap/{p}_1 \rlap/{p}_2
 + f_{35}g_{\mu\nu} \rlap/{p}_1\rlap/{p}_2
 + f_{36}g_{\mu\nu} \gamma_5\rlap/{p}_1 \rlap/{p}_2
 \}v(k_2)
\end{eqnarray*}
where ${\cal M}^{b},~{\cal M}^{q}$, and ${\cal M}^{tr}$ are the matrix elements contributed by
box, quartic and triangle interaction diagrams, respectively.
The cross section for this subprocess at one loop order in
unpolarized gluon collisions can be obtained by
\begin{equation}
\label{sss}
\hat{\sigma}(\hat{s},gg \rightarrow \tilde{\chi}_1^+ \mu^-) =
       \frac{1}{16 \pi \hat{s}^2} \int_{\hat{t}^{-}}^{\hat{t}^{+}}
       d\hat{t}~ \bar{\sum\limits_{}^{}} |{\cal M}|^2.
\end{equation}
In above equation, $\hat{t}$ is the momentum transfer squared from one
of the incoming gluons to the charged boson in the final state, and
$$
\hat{t}^\pm=\frac{1}{2}\left[ (m^{2}_{\tilde{\chi}_1^+}+m^{2}_\mu
-\hat{s})\pm \sqrt{(m^{2}_{\tilde{\chi}_1^+}+m^{2}_\mu-\hat{s})^2
-4m^{2}_{\tilde{\chi}_1^+}m^{2}_\mu}\right].
$$
The bar over the sum means average over initial spin and color.
\par
With the results from Eq.(\ref{sss}), we can easily obtain the total cross
section at $pp$ collider by folding the cross section of subprocess
$\hat{\sigma} (gg \rightarrow \tilde{\chi}_1^+ \mu^-)$ with the gluon
luminosity.

\begin{equation}
\sigma(s,pp \rightarrow gg \rightarrow  \tilde{\chi}_1^+\mu^-+X)=
   \int_{(m_{\tilde{\chi}_1^+}+m_{\mu})^2/
s} ^{1} d \tau \frac{d{\cal L}_{gg}}{d \tau} \hat{\sigma}
(gg\rightarrow \tilde{\chi}_1^+ \mu^-
\hskip 3mm at \hskip 3mm \hat{s}=\tau s),
\end{equation}

where $\sqrt{s}$ and $\sqrt{\hat{s}}$ are the $pp$ and $gg$
c.m.s. energies respectively and $d{\cal L}_{gg}/d \tau$
is the distribution function of gluon luminosity, which is defined as
\begin{equation}
\frac{d{\cal L}_{gg}}{d\tau}=\int_{\tau}^{1}
\frac{dx_1}{x_1} \left[ f_g(x_1,Q^2)f_g(\frac{\tau}{x_1},Q^2) \right].
\end{equation}
here $\tau = x_1~x_2$, the definition of $x_1$ and $x_2$ are from \cite{jiang},
and in our calculation we adopt the MRS set G parton distribution function
\cite{Martin}. The factorization scale Q was chosen as the average of the
final particles masses $\frac{1}{2}(m_{\tilde{\chi}_1^+} +m_{\mu})$.

\par
\section{Numerical results and discussions}
\par
In this section, we present some numerical results of the total cross
section from the complete one-loop diagrams for the
processes $pp \rightarrow gg \rightarrow \tilde{\chi}_1^{+}
\mu^{-}$. In our numerical calculation to get the low energy scenario
from the mSUGRA\cite{msugra}, the complete 2-loop renormalisation group
equations(RGE's) of the superpotential parameters for the supersymmetric
standard model including the full set of R-parity violating couplings
have been obtained in \cite{Allanach}, here we neglect the effects of
R-parity violation in the RGE's for simplicity. The RGE's (RGE's)\cite{RGE}
are run from the weak scale $m_Z$ up to the GUT scale, taking all
thresholds into account. We use two loop RGE's only for the gauge couplings
and the one-loop RGE's for the other supersymmetric parameters.
The GUT scale boundary conditions are imposed and the RGE's are run back to
$m_Z$, again taking threshold into account. We chose $R_{p}$-Parity
violating parameters concerned in the subprocess via gluon-gluon fusion to be
$\lambda_{211}^{'}=0.05$, $\lambda_{222}^{'}=0.21$ and $\lambda_{233}^{'}=0.3$,
which satisfy the constraints given by \cite{allanach2}.
The SM input parameters are chosen as: $m_t=173.8~GeV$, $m_{Z}=91.187~GeV$,
$m_b=4.5~GeV$, $\sin^2{\theta_{W}}=0.2315$, and $\alpha_{EW} = 1/128$. We
take a simple one-loop formula for the runing strong coupling constant
$\alpha_s$.

\begin{equation}
\alpha_s(\mu)=\frac{\alpha_s(m_Z)}{1+\frac{33-2n_f}{6\pi}\alpha_s(m_Z)
\ln{\frac{\mu}{m_Z}}}.
\end{equation}

where $\alpha_s(m_Z)=0.117$ and $n_f$ is the number of active flavors
at energy scale $\mu$.
\par
The cross sections for $\tilde{\chi}_1^{+}\mu^{-}$ via $gg$ collisions
at hadron colliders versus the mass of $\tilde{\chi}_1^{+}$ is shown in Fig.2.
The input parameters are chosen as $m_{0}=400~GeV$, $A_{0}= 300~GeV$,
$\tan{\beta}=4$, and $m_{1/2}$ varies from $150~GeV$ to $330~GeV$.
With above chosen parameters, we get that the value of $m_{\tilde{\chi}_1^+}$
varies from $104~GeV$ to $270~GeV$ in the framework of the mSUGRA as shown in
Fig.2. We calculate the cross sections at the LHC with the energies of
$\sqrt{s}$ being $14~TeV$. For the comparison, we also present the cross
section of $\tilde{\chi}_1^{+} \mu^{-}$ via quark-antiquark with the same
input parameters in Fig.2. It shows the cross section contribution to
parent process at hadron collider from subprocess $gg \rightarrow
\tilde{\chi}_1^{+} \mu^{-}$ can be competitive with
that via $q\bar{q}$ annihilation.
Then the production mechanism of subprocess $gg \rightarrow
\tilde{\chi}_1^{+}\mu^{-}$ should be considered in detecting the
$\rlap/{R}_{p}$ signals in this parameter space. We can see from Fig.2 that
the cross sections for $pp \rightarrow gg \rightarrow \tilde{\chi}_1^+ \mu^{-}$
decreases with the increment of the mass of $\tilde{\chi}_1^+$. It can reach
10.3 femto barn when $m_{\tilde{\chi}_1^{+}}$ is about $104~GeV$.
\par
In Fig.3 we present the cross sections of $\tilde{\chi}_1^{+}\mu^{-}$
productions versus the mass of $\mu$ sneutrino with the collision energy
of hadron being $14~TeV$, where the input parameters
are chosen as  $m_{1/2}=150~GeV$, $A_{0}= 300~GeV$ and $\tan{\beta}=4$.
In the mSUGRA scenario, when $m_{0}$ increases from $200~GeV$ to $800~GeV$,
the mass of $\mu$ sneutrino ranges from $219~GeV$ to $804~GeV$. We can see
from Fig.3 that the cross section decreases rapidly with the increment of
the mass of $\mu$ sneutrino, this is because the cross section is enhanced
by the $\mu$ sneutrino resonance effects in the lower $\mu$ sneutrino mass
range. We also show the cross section contributed by the $\hat{s}$ channel
diagrams. It shows that the cross section are mostly contributed by the effects
of resonance, only when the mass of $\mu$ sneutrino is larger than $700~GeV$,
the non-resonant contributions can reach about $15\%$ to the total
cross sections.
\par
The cross sections for the production of single chargino accompanied by $\mu$
lepton via gluon-gluon versus $\sqrt{s}$ with $m_{1/2}=150$ GeV, $m_0=400$
GeV, $A_0=300$ GeV, $\tan\beta$=4 are depicted in Fig.4. The solid line is for
$\mu>0$, and the dashed line is for $\mu<0$. The discrepancy between these two
curves is not very large. This feature can be seen also in the corresponding
curves in Fig.2 and Fig.3. It shows that the production rate of process
$pp \rightarrow gg \rightarrow \tilde{\chi}_1^{\pm} \mu^{\mp}+X$ is not very
sensitive to the sign of parameter $\mu$.

\par
Finally, we will discuss the relationship between the cross section and the
parameter $\lambda_{ijk}'$. The cross section of $\tilde{\chi}_1^+ \mu^-$
production via $q\bar{q}$ annihilation is scaled by $\lambda_{211}^{'2}$.
While the cross section via $gg$ should take the sum of three generation of
(s)quarks loop, but the main contribution to the cross section is from the
third generation for the coupling coefficient of the third generation is much
larger than the first and second generation, then the cross section is nearly
proportional to the $\lambda_{233}^{'2}$. If $\lambda_{233}^{'}$ is less than
0.18, The cross section via $gg$ will be less than 3.5 femto barn, in this case,
compared with the process of $q\bar{q}$ annihilation, it can be neglected.
But if $\lambda_{233}^{'}$ is larger than the value we discussed, the
process via $gg$ will play a more important rule in $pp$ collision. In fact,
the maximal possible cross section could be somewhat higher than indicated
in fig.3, because the upper bounds of R-parity violation couplings can be higher
for heavy sparticls.
\par
\section{Summary}
\par
In this paper, we have studied the production of single chargino
$\tilde{\chi}_1^{\pm}$ associated with $\mu$ lepton with explicit
$R_{p}$-violation at hadron colliders. The production rates via gluon-gluon
fusion at the LHC are numerically analysed in the mSUGRA scenario with some
typical parameter sets. The results show that the cross section of the
$\tilde{\chi}_1^{\pm}\mu^{\mp}$ production via gluon-gluon collisions can
reach about some femto barn to hundreds femto barn at the LHC with
our chosen parameters. It shows that the production mechanism via gluon-gluon
fusion can be competitive with that from quark-antiquark annihilation process.
Therefore, In detecting the
$\tilde{\chi}_1^{\pm} \mu^{\mp}$ productions at the LHC to search for the
signals of both SUSY and $R_{p}$ violation, we should consider not only the
$\tilde{\chi}_1^{\pm} \mu^{\mp}$ production subprocesses via quark-antiquark
annihilation, but also those via the gluon-gluon fusion.

\vskip 4mm
\noindent{\large\bf Acknowledgement:}
This work was supported in part by the National Natural Science
Foundation of China(project number: 19875049), the Youth Science
Foundation of the University of Science and Technology of China, a grant
from the Education Ministry of China and the State Commission of Science
and Technology of China.

\par
One of the author Wan Lang-Hui would thank Zhou Mian-Lai, Zhou Fei and
Zhou Hong for useful discussion.

\par
\renewcommand{\theequation}{A.\arabic{equation}}
\begin{flushleft} {\bf Appendix} \end{flushleft}
\par
The Feynman rules for the couplings we used are list below:
\begin{eqnarray*}
\bar{U}_n-\tilde{D}_{n,i}-\tilde{\chi}_j^+ :~~~~V_{U_n \tilde{D}_{n,i}
      \tilde{\chi}_j^+}^{(1)}P_L +
       V_{U_n \tilde{D}_{n,i} \tilde{\chi}_j^+}^{(2)}P_R,\\
\bar{D}_n-\tilde{U}_{n,i}-\tilde{\chi}_j^{+c} :~~~~\{ V_{D_n \tilde{U}_{n,i}
      \tilde{\chi}_j^+}^{(1)}P_L +
       V_{D_n \tilde{U}_{n,i} \tilde{\chi}_j^+}^{(2)}P_R \} C,\\
\bar{E}_n-\tilde{\nu}_{E_n}-\tilde{\chi}_j^{+c} :~~~~\{ V_{E_n \tilde{\nu}_{E_n}
      \tilde{\chi}_j^+}^{(1)}P_L +
       V_{E_n \tilde{\nu}_{E_n} \tilde{\chi}_j^+}^{(2)}P_R \}C,\\
\tilde{\nu}_i-D_j-\bar{D}_k:~~~~V_{\tilde{\nu}_iD_jD_k}^{(1)} P_L +
                  V_{\tilde{\nu}_iD_jD_k}^{(2)} P_R\\
\bar{E}_i-D_j-\tilde{U}_{k,n}:~~~~ V_{E_iD_j\tilde{U}_{k,n}}^{(2)}P_R\\
E_i^c-\bar{U}_j-\tilde{D}_{k,n}:~~~~ V_{E_iU_j\tilde{D}_{k,n}}^{(2)}P_R~C\\
\tilde{\nu_i}-\tilde{D}_{j,l}-\bar{\tilde{D}}_{k,n}:~~~~
               V_{\tilde{\nu}_{i} \tilde{D}_{j,l} \tilde{D}_{k,n}}
\end{eqnarray*}
\par

Where $C$ ic the charge conjugation operator, $P_{L,R} = \frac{1}{2}(1 \mp
\gamma_5)$. The lower-index in $Q_i(E_i)$ refers to the generation index
of quarks and leptons, the lower-indices in $\tilde{Q}_{i,j}(Q=U,D)$ represent
generation index and index of the physical squark, respectively.
For the expressions of $V_{U_n \tilde{D}_{n,i} \tilde{\chi}_j^+}^{(1,2)}$
and $V_{D_n \tilde{U}_{n,i} \tilde{\chi}_j^+}^{(1,2)}$, one can refer Ref.\cite{Mwg},
the forms of $V_{E_n \tilde{\nu}_{E_n} \tilde{\chi}_j^+}^{(1,2)}$ can be find in
Ref.\cite{r1}. The other vertices can be read out from Eq.(\ref{lag}):
$$
V_{\tilde{\nu}_iD_jD_k}^{(1)}=-i\lambda_{ijk}^{'},~~~~~~~~~
V_{\tilde{\nu}_iD_jD_k}^{(2)}=-i\lambda_{ikj}^{'}
$$
$$
V_{E_iD_j\tilde{U}_{k,1}}^{(2)}=i\lambda_{ijk}^{'}\cos \theta_{U_k},~~~
V_{E_iD_j\tilde{U}_{k,2}}^{(2)}=i\lambda_{ijk}^{'}\sin \theta_{U_k},
$$
$$
V_{E_iU_j\tilde{D}_{k,1}}^{(2)} =-i\lambda_{ijk}^{'}\sin \theta_{D_k},~~~
V_{E_iU_j\tilde{D}_{k,2}}^{(2)} =i\lambda_{ijk}^{'}\cos \theta_{D_k}.
$$
$$
V_{\tilde{\nu}_{i} \tilde{D}_{j,1} \tilde{D}_{k,1}} = -i \lambda_{ijk}^{'}
      \left[ m_{D_j}  \sin \theta_{D_j} \sin \theta_{D_k}
      +m_{D_k} \cos \theta_{D_j} \cos \theta_{D_k}  -
      A_{d} \cos \theta_{D_{j}} \sin \theta_{D_{k}} \right].
$$
$$
V_{\tilde{\nu}_{i} \tilde{D}_{j,2} \tilde{D}_{k,2}} = -i \lambda_{ijk}^{'}
      \left[ m_{D_j}  \cos \theta_{D_j} \cos \theta_{D_k}
      +m_{D_k} \sin \theta_{D_j} \sin \theta_{D_k}  +
      A_{d} \sin \theta_{D_{j}} \cos \theta_{D_{k}}  \right].
$$
\par
where $A_d$ is the soft breaking parameter.
The amplitude parts for the $u$-channel box and triangle vertex interaction
diagrams can be obtained from the $t$-channel's by doing exchanges as shown
below:
\begin{equation}
{\cal M}^{\hat{u}} = {\cal\ M}^{\hat t}(\hat{t}\rightarrow \hat{u}, k_1
              \leftrightarrow k_2, \mu \leftrightarrow \nu),
\end{equation}
\par
So we present only the $t$-channel form factors for box and triangle diagrams.
The form factors for the figures with loop of $U$ quark and $\tilde{D}$ squark
in Fig.1(a,b), can be obtained from the form factors corresponding to the
figures with loop of $D$ quark and $\tilde{U}$ squark in Fig.1(a.1), (b.1) by
doing the replacement of $m_D \rightarrow m_U$, $m_{\tilde{U}_{j,k}}
\rightarrow m_{\tilde{D}_{j,k}}$ , $F_{1,k} \rightarrow F_{3,k},
~F_{2,k} \rightarrow F_{4,k}$, $C^{1,k} \rightarrow C^{3,k}$, $D^{1,k}
\rightarrow D^{4,k}$,$D^{2,k} \rightarrow D^{5,k}$ and $D^{3,k} \rightarrow
D^{6,k}$. Since the form factors of figures with the first or second
generation quark and squark loop are analogous to corresponding ones with
the third generation quark and squark loop, here we give only the form
factors for the figures in Fig.1(a),(b) including the third generation quark
and squark loop. In this appendix, we use the notations defined below for
abbreviation:
\begin{eqnarray*}
&&B_{0}=B_{0}[-p1, m_b, m_b]\\
&&B_{0}^1=B_{0}[\hat{s},m_{\tilde{b}_k},m_{\tilde{b}_k}]\\
&&C_{0}^{1,k},C_{ij}^{1,k}=C_{0},C_{ij}[
  k_2, k_1, m_{\tilde{t}_k}, m_b, m_{\tilde{t}_k}]\\
&&C_{0}^{2}=C_{0}[
    -p_1, -p_2, m_b, m_b, m_b]\\
&&C_{0}^{3,k},C_{ij}^{3,k}=C_{0},C_{ij}[
  k_2, k_1, m_{\tilde{b}_k}, m_t, m_{\tilde{b}_k}]\\
&&C_{0}^{4},C_{ij}^{4} = C_{0},C_{ij}[-p_2, -p_1, m_b, m_b, m_b]\\
&&C_{0}^{5,k},C_{ij}^{5,k} = C_{0},C_{ij}[-p_2, -p_1, m_{\tilde{b}_k},
                                 m_{\tilde{b}_k},m_{\tilde{b}_k}]\\
&&D_{0}^{1,k},D_{i,j}^{1,k},D_{ijl}^{1,k}=D_{0},D_{i,j},D_{ijl}[
   -p_2, k_1, -p_1, m_{\tilde{t}_k},m_{\tilde{t}_k}, m_b, m_b]\\
&&D_{0}^{2,k},D_{ij}^{2,k},D_{ijl}^{2,k}=D_{0},D_{ij},D_{ijl}[
    k_1, -p_1, -p_2, m_{\tilde{t}_k},m_b,m_b,m_b]\\
&&D_{0}^{3,k},D_{ij}^{3,k},D_{ijl}^{3,k}=D_{0},D_{ij},D_{ijl}[
   k_1, -p_1, -p_2, m_b, m_{\tilde{t}_k},m_{\tilde{t}_k}, m_{\tilde{t}_k}]\\
&&D_{0}^{4,k},D_{ij}^{4,k},D_{ijl}^{4,k}=D_{0},D_{ij},D_{ijl}[
   -p_2, k_1, -p_1, m_{\tilde{b}_k}, m_{\tilde{b}_k}, m_t, m_t]\\
&&D_{0}^{5,k},D_{ij}^{5,k},D_{ijl}^{5,k}=D_{0},D_{ij},D_{ijl}[
   k_1, -p_1, -p_2, m_{\tilde{b}_k}, m_t, m_t, m_t]\\
&&D_{0}^{6,k},D_{ij}^{6,k},D_{ijl}^{6,k}=D_{0},D_{ij},D_{ijl}[
   k_1, -p_1, -p_2, m_t, m_{\tilde{b}_k}, m_{\tilde{b}_k}, m_{\tilde{b}_k}]\\
&&A_{\tilde{\nu}} = \frac{i}{\hat{s}-m_{\tilde{\nu}_{\mu}}^2+
                    i m_{\tilde{\nu}_{\mu}} \Gamma_{\tilde{\nu}_{\mu}}  }\\
&&F_{1,k} = V_{\mu b\tilde{t}_{k}}^{(2)}V_{b \tilde{t}_k \tilde{\chi}_1^+}^{(1)}\\
&&F_{2,k} = V_{\mu b\tilde{t}_{k}}^{(2)}V_{b \tilde{t}_k \tilde{\chi}_1^+}^{(2)}\\
&&F_{3,k} = V_{t \tilde{b}_k \tilde{\chi}_1^+}^{(2)*}V_{\mu t \tilde{b}_{k}}^{(2)} \\
&&F_{4,k} = V_{t \tilde{b}_k \tilde{\chi}_1^+}^{(1)*}V_{\mu t \tilde{b}_{k}}^{(2)}
\end{eqnarray*}
\par
The form factors in the amplitude of the quartic
interaction diagrams Fig.1(b) and  are expressed as
\begin{eqnarray*}
f_1^q&=&
\frac{ig_{s}^2}{32\pi^2} \sum_{k=1}^{2}
    (C_{0}^{1,k} F_{2,k} m_b- C_{12}^{1,k} F_{1,k} m_{\tilde{\chi}_1^+}
    +(C_{0}^{1,k}+C_{11}^{1,k}) F_{1,k} m_{\mu}
    + B_0^1 V_{\tilde{\nu}_{\mu} \tilde{b}_k \tilde{b}_k }^{*} A_{\tilde{\nu}}
    V_{\mu \tilde{\nu}_{\mu} \tilde{\chi}_1^+}^{(2)} )  \\
f_2^q&=&
\frac{ig_{s}^2}{32\pi^2} \sum_{k=1}^{2}
     (C_{0}^{1,k} F_{2,k} m_b- C_{12}^{1,k} F_{1,k} m_{\tilde{\chi}_1^+}
     -(C_{0}^{1,k}+C_{11}^{1,k}) F_{1,k} m_{\mu}
    + B_0^1 V_{\tilde{\nu}_{\mu} \tilde{b}_k \tilde{b}_k }^{*} A_{\tilde{\nu}}
    V_{\mu \tilde{\nu}_{\mu} \tilde{\chi}_1^+}^{(2)} )  \\
f_{i}^q&=&0~~~(i=3-36)
\end{eqnarray*}
\par
The form factors in the amplitude from the $t$-channel triangle diagrams depicted in fig.1(c)
are list below:
\begin{eqnarray*}
f_1^{tr}&=&f_2^{tr}=
     -\frac{g_S^2}{64 \pi^2}
     \left( 2 B_0 - 8 C_{24}^4 - C_0^4 - 2 C_{12}^4 s  \right) m_b A_{\tilde{\nu}}
     \left( V_{\tilde{\nu}_{\mu}bb}^{(1)*} +V_{\tilde{\nu}_{\mu}bb}^{(2)*} \right)
      V_{\mu \tilde{\nu}_{\mu} \tilde{\chi}_1^+}^{(2)} \\
   &-&\frac{ig_s^2}{16 \pi^2}\sum_{k=1}^2
   C_{24}^{5,k} V_{\tilde{\nu}_{\mu} \tilde{b}_k \tilde{b}_k }^{*} A_{\tilde{\nu}}
     V_{\mu \tilde{\nu}_{\mu} \tilde{\chi}_1^+}^{(2)}\\
f_{i}^{tr}&=&0~~~(i=3-36)
\end{eqnarray*}
\par
The form factors of the amplitude part from $t$-channel box diagrams, Fig.1 (a) are written as
\begin{eqnarray*}
f_1^{b,t}&=&f_{2}^{b,t}=
    -\frac{ig_{s}^2}{16\pi^2}\sum_{k=1}^{2}
    \left[ (D_{27}^{1,k}+D_{27}^{2,k}+D_{27}^{3,k}) F_{2,k} m_b \right.\\
    &+& (D_{27}^{1,k}+ D_{312}^{1,k} +D_{27}^{2,k}
    +\left. D_{311}^{2,k} -D_{311}^{3,k} ) F_{1,k} m_{\tilde{\chi}_1^+}\right]\\
f_3^{b,t}&=&f_{4}^{b,t}=
    \frac{ig_{s}^2}{16\pi^2}\sum_{k=1}^{2}
     \left[ (D_{12}^{1,k}+D_{22}^{1,k}+ D_{0}^{2,k}
     + 2 D_{11}^{2,k} + D_{21}^{2,k} + D_0^{3,k} + 2 D_{11}^{3,k}
     + D_{21}^{3,k}) F_{2,k} m_b \right.\\
     &+& \left. (D_{12}^{1,k}+2 D_{22}^{1,k}
     + D_{32}^{1,k}+D_{0}^{2,k}+3 D_{11}^{2,k}+3 D_{21}^{2,k} +  D_{31}^{2,k}
     - D_{11}^{3,k}-2 D_{21}^{3,k}-D_{31}^{3,k}) F_{1,k} m_{\tilde{\chi}_1^+}\right]\\
f_5^{b,t}&=&
     \frac{ig_{s}^2}{32\pi^2}\sum_{k=1}^{2}
     F_{1,k} \left[2 D_{27}^{1,k}+4 D_{312}^{1,k}
     - D_{12}^{1,k} \hat{s}-D_{24}^{1,k} \hat{s}
     - D_{26}^{1,k} \hat{s}-D_{310}^{1,k} \hat{s}-D_{26}^{1,k} \hat{t} \right.\\
     &-& D_{38}^{1,k} \hat{t}+ 4 D_{27}^{2,k}
     + 4 D_{311}^{2,k}
     - D_{13}^{2,k} \hat{s}-D_{25}^{2,k} \hat{s}
     - D_{26}^{2,k} \hat{s}-D_{310}^{2,k} \hat{s}-D_{12}^{2,k} \hat{t}
     - 2D_{24}^{2,k} \hat{t} \\
     &-& D_{12}^{1,k} \hat{u}-D_{22}^{1,k} \hat{u} - D_{24}^{1,k} \hat{u}
     - D_{36}^{1,k} \hat{u}+D_{12}^{1,k} m_b^2- D_{34}^{2,k} \hat{t}-D_{13}^{2,k} \hat{u}
     - 2 D_{25}^{2,k} \hat{u} - D_{35}^{2,k} \hat{u}\\
     &+& (D_{0}^{2,k}
     + D_{11}^{2,k}) m_b^2+2D_{27}^{3,k}
     + 2D_{311}^{3,k}+(D_{13}^{2,k}- D_{0}^{2,k}-3 D_{11}^{2,k}+D_{12}^{2,k}
     -  3 D_{21}^{2,k}+2 D_{24}^{2,k}\\
     &+& 2 D_{25}^{2,k}
     -  D_{31}^{2,k}+D_{34}^{2,k}+D_{35}^{2,k}
     + \left. D_{24}^{1,k} -D_{22}^{1,k}
     + D_{26}^{1,k} -D_{32}^{1,k}
     + D_{36}^{1,k} +D_{38}^{1,k}) m_{\tilde{\chi}_1^+}^2\right]\\
f_6^{b,t}&=&
     - \frac{ig_{s}^2}{32\pi^2}\sum_{k=1}^{2}
     F_{1,k} \left[ 2D_{27}^{1,k}+2D_{312}^{1,k}
     + C_{0}^{2}-2 D_{27}^{2,k}+2D_{311}^{2,k}+D_{26}^{2,k} \hat{s}+D_{24}^{2,k} \hat{t}
     + D_{25}^{2,k} \hat{u} \right.\\
     &-& 2D_{27}^{3,k}-2D_{311}^{3,k}+ \left.( D_{21}^{2,k}-D_{24}^{2,k}-D_{25}^{2,k})
     m_{\tilde{\chi}_1^+}^2-D_{0}^{2,k} m_{\tilde{t}_k}^2 \right]\\
f_7^{b,t}&=&
     \frac{ig_{s}^2}{16\pi^2} \sum_{k=1}^{2}
     F_{1,k}\left[ D_{27}^{1,k}+D_{313}^{1,k}
     + D_{27}^{2,k}+D_{312}^{2,k}-D_{312}^{3,k}\right]\\
f_8^{b,t}&=&
     -\frac{ig_{s}^2}{16\pi^2} \sum_{k=1}^{2}
     F_{1,k}\left[ D_{26}^{1,k}+D_{38}^{1,k}
     + D_{12}^{2,k}+2 D_{24}^{2,k}+D_{34}^{2,k}-D_{12}^{3,k}-2 D_{24}^{3,k}-D_{34}^{3,k}\right]\\
f_9^{b,t}&=&
    \frac{ig_{s}^2}{16\pi^2}\sum_{k=1}^{2}
    F_{1,k} \left[ D_{27}^{1,k}+D_{311}^{1,k}
    + D_{313}^{2,k}-D_{313}^{3,k} \right]\\
f_{10}^{b,t}&=&
    -\frac{ig_{s}^2}{16\pi^2}\sum_{k=1}^{2}
    F_{1,k}\left[ D_{12}^{1,k}+D_{22}^{1,k}+D_{24}^{1,k}+D_{36}^{1,k}
    + D_{13}^{2,k}+2 D_{25}^{2,k}+D_{35}^{2,k} \right.\\
    &-& \left. D_{13}^{3,k}-2 D_{25}^{3,k}-D_{35}^{3,k}\right]\\
f_{11}^{b,t}&=&
    -\frac{ig_{s}^2}{32\pi^2} F_{1,k}\sum_{k=1}^{2}
     \left[ 2 D_{27}^{1,k}+4 D_{312}^{1,k}
    - D_{12}^{1,k} \hat{s}-D_{24}^{1,k} \hat{s}-D_{26}^{1,k} \hat{s}-D_{310}^{1,k} \hat{s}
    - D_{26}^{1,k} \hat{t}- D_{38}^{1,k} \hat{t} \right.\\
    &-& D_{12}^{1,k} \hat{u}
    -D_{22}^{1,k} \hat{u}-D_{24}^{1,k} \hat{u}-D_{36}^{1,k}\hat{u}
    + 4 D_{27}^{2,k}+4 D_{311}^{2,k}- D_{13}^{2,k} \hat{s}-D_{25}^{2,k} \hat{s}
    - D_{26}^{2,k} \hat{s}-D_{310}^{2,k} \hat{s} \\
    &-& D_{12}^{2,k} \hat{t}
    - 2 D_{24}^{2,k} \hat{t}
    - D_{34}^{2,k} \hat{t}-D_{13}^{2,k} \hat{u}-2 D_{25}^{2,k} \hat{u}- D_{35}^{2,k} \hat{u}
    + 2D_{27}^{3,k}+2D_{311}^{3,k}+(D_{12}^{1,k}\\
    &+& D_{0}^{2,k} + D_{11}^{2,k}) m_b^2
    + (D_{24}^{1,k} -D_{22}^{1,k}+D_{26}^{1,k}-D_{32}^{1,k}+ D_{36}^{1,k}
    + D_{38}^{1,k}+D_{13}^{2,k}\\
    &-& D_{0}^{2,k}-3 D_{11}^{2,k}
    + D_{12}^{2,k}-3 D_{21}^{2,k}+2 D_{24}^{2,k}+2 D_{25}^{2,k}
    - \left. D_{31}^{2,k}+D_{34}^{2,k}+D_{35}^{2,k}) m_{\tilde{\chi}_1^+}^2  \right]\\
f_{12}^{b,t}&=&
    \frac{ig_{s}^2}{32\pi^2} \sum_{k=1}^{2}
    F_{1,k}\left[ 2D_{27}^{1,k}+2D_{312}^{1,k}
    + C_{0}^{2}-2 D_{27}^{2,k}+2D_{311}^{2,k}+D_{26}^{2,k} \hat{s}
    +D_{24}^{2,k} \hat{t} \right.\\
    &+& \left. D_{25}^{2,k} \hat{u}-2D_{27}^{3,k} - 2D_{311}^{3,k}+(D_{21}^{2,k}
    - D_{24}^{2,k} -D_{25}^{2,k})m_{\tilde{\chi}_1^+}^2-D_{0}^{2,k} m_{\tilde{t}_k}^2 \right]\\
f_{13}^{b,t}&=&
    -\frac{ig_{s}^2}{16\pi^2}\sum_{k=1}^{2}
    F_{1,k} \left[ D_{27}^{1,k}+D_{313}^{1,k}
    + D_{27}^{2,k}+D_{312}^{2,k}-D_{312}^{3,k} \right]\\
f_{14}^{b,t}&=&
    \frac{ig_{s}^2}{16\pi^2}\sum_{k=1}^{2}
    F_{1,k} \left[ D_{26}^{1,k}+D_{38}^{1,k}
    + D_{12}^{2,k}+2 D_{24}^{2,k}+D_{34}^{2,k}
    - D_{12}^{3,k}-2 D_{24}^{3,k}-D_{34}^{3,k} \right]\\
f_{15}^{b,t}&=&
    -\frac{ig_{s}^2}{16\pi^2}\sum_{k=1}^{2}
    F_{1,k}\left[ D_{27}^{1,k}+D_{311}^{1,k}
    + D_{313}^{2,k}-D_{313}^{3,k} \right]\\
f_{16}^{b,t}&=&
    \frac{ig_{s}^2}{16\pi^2}\sum_{k=1}^{2}
    F_{1,k} \left[ D_{12}^{1,k}+D_{22}^{1,k}+D_{24}^{1,k}+D_{36}^{1,k}
    + D_{13}^{2,k}+2 D_{25}^{2,k}+D_{35}^{2,k} \right.\\
    &-& \left. D_{13}^{3,k}-2 D_{25}^{3,k}-D_{35}^{3,k}\right]\\
f_{17}^{b,t}&=&f_{18}^{b,t}=
    -\frac{ig_{s}^2}{32\pi^2}\sum_{k=1}^{2}
    \left[ (D_{12}^{1,k}+D_{0}^{2,k}+D_{11}^{2,k}) F_{2,k} m_b
    + (D_{12}^{1,k}+D_{22}^{1,k}
    + D_{0}^{2,k}+2 D_{11}^{2,k} \right.\\
    &+& \left. D_{21}^{2,k}) F_{1,k} m_{\tilde{\chi}_1^+} \right]\\
f_{19}^{b,t}&=&-f_{21}^{b,t}=
    -\frac{ig_{s}^2}{64\pi^2}\sum_{k=1}^{2}
    F_{1,k} \left[2D_{27}^{1,k}+ 4 D_{27}^{2,k}+6 D_{312}^{2,k}
    - D_{26}^{2,k} \hat{s}-D_{38}^{2,k} \hat{s}-D_{12}^{2,k} \hat{t} \right.\\
    &-& D_{22}^{2,k} \hat{t}-D_{24}^{2,k} \hat{t}
    - D_{36}^{2,k} \hat{t}
    - D_{26}^{2,k} \hat{u}-D_{310}^{2,k} \hat{u}+D_{12}^{2,k} m_b^2+(D_{22}^{2,k}
    - D_{24}^{2,k}+D_{26}^{2,k}\\
    &+& \left. D_{310}^{2,k}-D_{34}^{2,k}+D_{36}^{2,k}) m_{\tilde{\chi}_1^+}^2 \right]\\
f_{20}^{b,t}&=&-f_{22}^{b,t}=
    \frac{ig_{s}^2}{32\pi^2}\sum_{k=1}^{2}
    F_{1,k} \left[ D_{12}^{1,k}+D_{24}^{1,k}+D_{13}^{2,k}+D_{25}^{2,k}\right]\\
f_{23}^{b,t}&=&f_{24}^{b,t}=
    -\frac{ig_{s}^2}{32\pi^2}\sum_{k=1}^{2}
     \left[ (D_{0}^{2,k}+D_{11}^{2,k}) F_{2,k} m_b
     +( D_{0}^{2,k}+2 D_{11}^{2,k}+D_{21}^{2,k}) F_{1,k} m_{\tilde{\chi}_1^+} \right] \\
f_{25}^{b,t}&=&-f_{27}^{b,t}=
    -\frac{ig_{s}^2}{64\pi^2}\sum_{k=1}^{2}
     F_{1,k} \left[ 2 D_{27}^{2,k}+6 D_{313}^{2,k}
     - D_{13}^{2,k} \hat{s}-D_{23}^{2,k}\hat{s}-D_{26}^{2,k} \hat{s}
     - D_{39}^{2,k} \hat{s} \right.\\
     &-& D_{12}^{2,k} \hat{t}-D_{13}^{2,k} \hat{t}
    - D_{24}^{2,k} \hat{t}-D_{25}^{2,k} \hat{t}-D_{26}^{2,k} \hat{t}
    - D_{310}^{2,k} \hat{t}-D_{13}^{2,k}\hat{u}-D_{23}^{2,k} \hat{u}
    - D_{25}^{2,k} \hat{u}-D_{37}^{2,k} \hat{u} \\
    &+& ( D_{0}^{2,k}+D_{13}^{2,k}) m_b^2
    + (D_{37}^{2,k}-D_{0}^{2,k}-2 D_{11}^{2,k}+D_{12}^{2,k}+D_{13}^{2,k}-
    D_{21}^{2,k}+D_{23}^{2,k}+D_{24}^{2,k}\\
    &+& \left. D_{26}^{2,k}+D_{310}^{2,k}-D_{35}^{2,k}) m_{\tilde{\chi}_1^+}^2 \right]  \\
f_{26}^{b,t}&=&-f_{28}^{b,t}=
    -\frac{ig_{s}^2}{32\pi^2}\sum_{k=1}^{2}
    F_{1,k}\left[ D_{12}^{2,k}+D_{24}^{2,k} \right]     \\
f_{31}^{b,t}&=&f_{32}^{b,t}=
    \frac{ig_{s}^2}{64\pi^2}\sum_{k=1}^{2}
    \left[ (4 D_{27}^{2,k}-D_{13}^{2,k} \hat{s}
    - D_{26}^{2,k} \hat{s}-D_{0}^{2,k} \hat{t}-D_{11}^{2,k}\hat{t}
    - D_{12}^{2,k} \hat{t}-D_{24}^{2,k} \hat{t}-D_{13}^{2,k} \hat{u} \right. \\
    &-& D_{25}^{2,k} \hat{u})F_{2,k} m_b
    + D_{0}^{2,k} F_{2,k} m_b^3+(6 D_{27}^{2,k}
    + 6D_{311}^{2,k} - D_{13}^{2,k} \hat{s}-D_{25}^{2,k} \hat{s}
    -D_{26}^{2,k} \hat{s} - D_{310}^{2,k} \hat{s} \\
    &-& D_{0}^{2,k}\hat{t}
    - 2 D_{11}^{2,k} \hat{t}
    - D_{12}^{2,k} \hat{t}-D_{21}^{2,k} \hat{t}-2D_{24}^{2,k} \hat{t}
    - D_{34}^{2,k} \hat{t}-D_{13}^{2,k} \hat{u}-2 D_{25}^{2,k} \hat{u}
    - D_{35}^{2,k} \hat{u} \\
    &+& D_{0}^{2,k} m_b^2
    + D_{11}^{2,k} m_b^2) F_{1,k} m_{\tilde{\chi}_1^+}
    +(D_{12}^{2,k}- D_{11}^{2,k}+D_{13}^{2,k}-D_{21}^{2,k}
    + D_{24}^{2,k}+D_{25}^{2,k}) F_{2,k} m_b m_{\tilde{\chi}_1^+}^2 \\
    &+&( D_{12}^{2,k}-D_{11}^{2,k}
    + D_{13}^{2,k}-2 D_{21}^{2,k}
    + \left. 2 D_{24}^{2,k}+2 D_{25}^{2,k}-D_{31}^{2,k}
    + D_{34}^{2,k}+D_{35}^{2,k}) F_{1,k} m_{\tilde{\chi}_1^+}^3 \right]                            \\
f_{33}^{b,t}&=& f_{34}^{b,t}=
    -\frac{ig_{s}^2}{64\pi^2}\sum_{k=1}^{2}
    \left[ D_{0}^{2,k} F_{2,k} m_b
    + (D_{0}^{2,k} + D_{11}^{2,k}) F_{1,k}  m_{\tilde{\chi}_1^+} \right]      \\
f_i^{b,t}&=&0~~~(i=29,30,35,36)
\end{eqnarray*}
\par
  In this work we adopted the definitions of two-, three-, and four-point one
loop Passarino-Veltman integral functions as shown in Ref.\cite{bernd} and
all the vector and tensor integrals can be deduced in the forms of scalar
integrals \cite{veltman}.

\vskip 10mm
\begin{flushleft} {\bf Figure Captions} \end{flushleft}

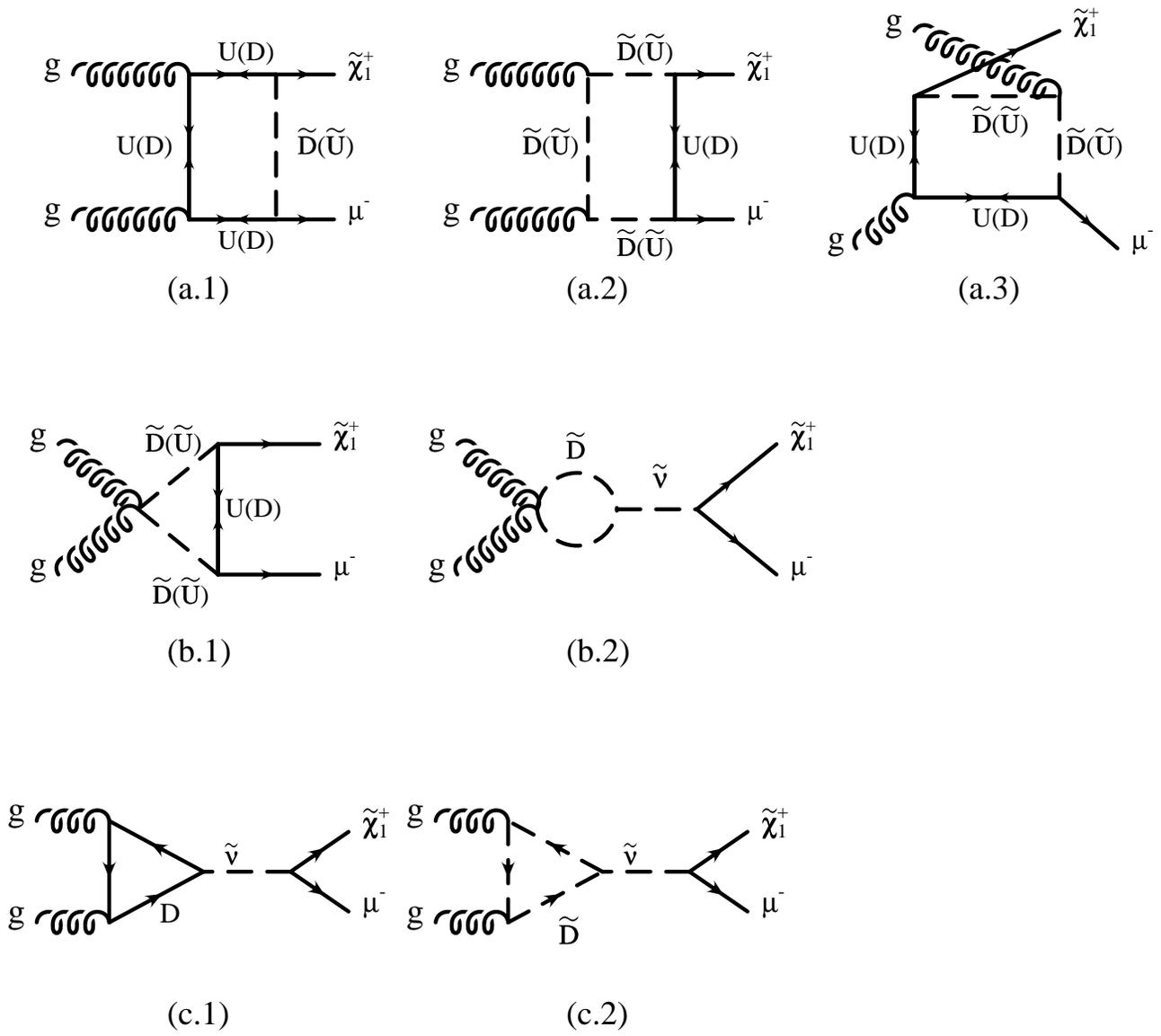
{\bf Fig.1} The Feynman diagrams of the subprocess $gg \rightarrow
\tilde{\chi}_1^{+}\mu^{-}$. ($a.1 \sim 3$) box diagrams.
($b.1$) quartic interaction diagrams. ($c.1$) triangle diagrams.

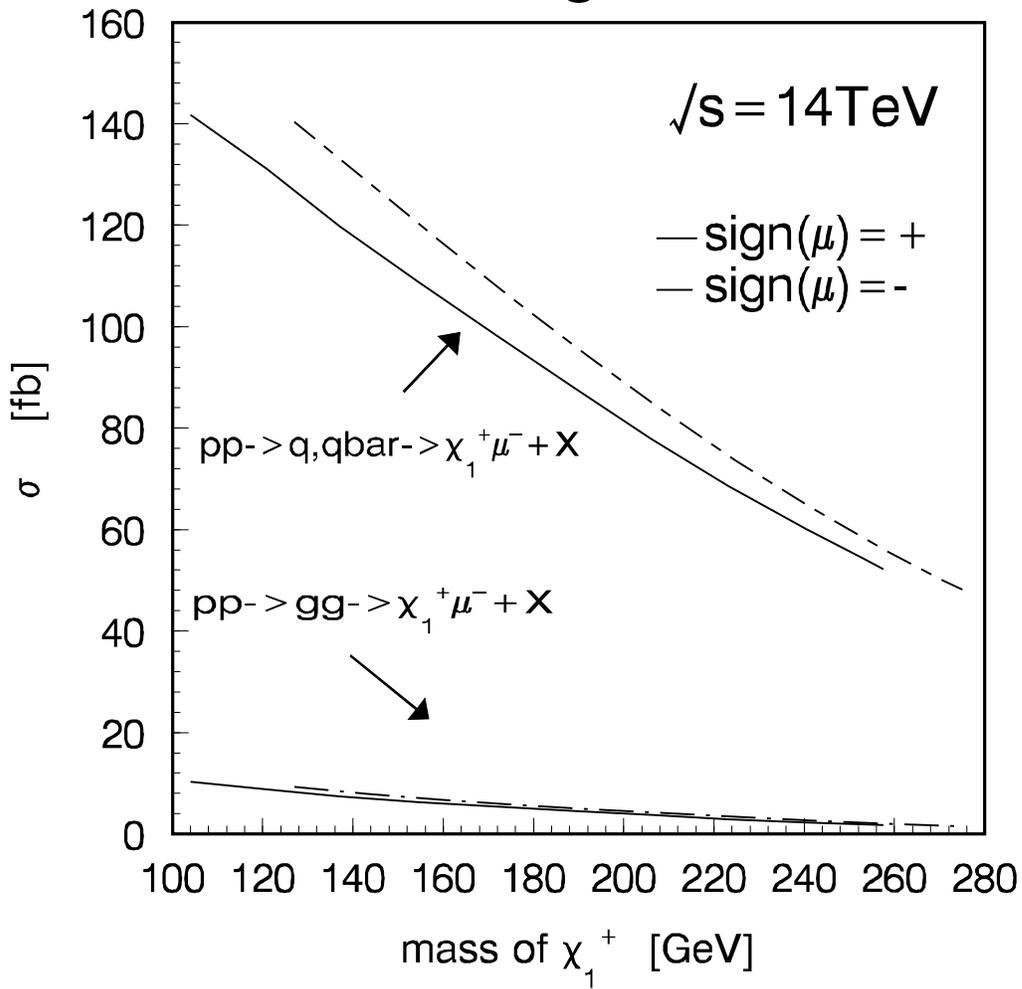
{\bf Fig.2} Total cross section of the process $pp \rightarrow gg \rightarrow
\tilde{\chi}_1^{+}\mu^{-}+X$ as a function of $m_{\tilde{\chi}_1^+}$ at the LHC
with $\sqrt{s}=14~TeV$.

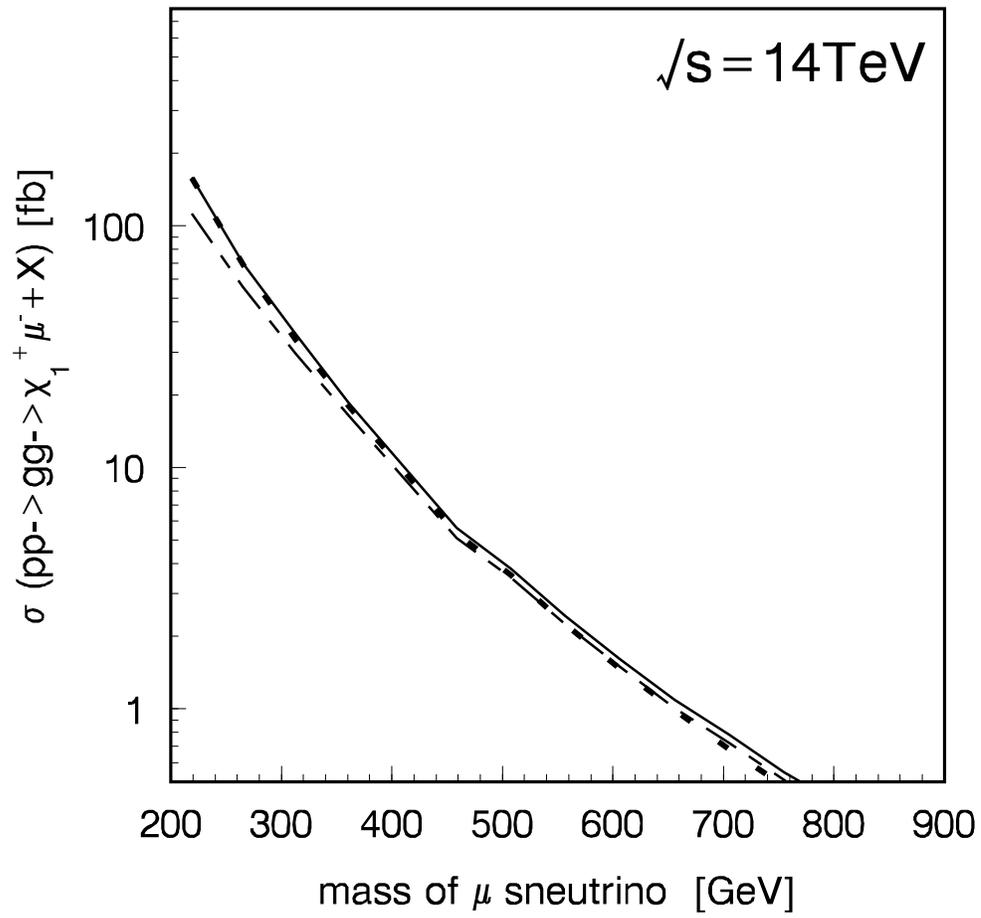
{\bf Fig.3} Total cross section of the process $pp \rightarrow
gg \rightarrow \tilde{\chi}_1^{+}\mu^{-}+X$ as a function of $m_{\tilde{\nu}_\mu}$
at the LHC with $\sqrt{s}=14~TeV$.

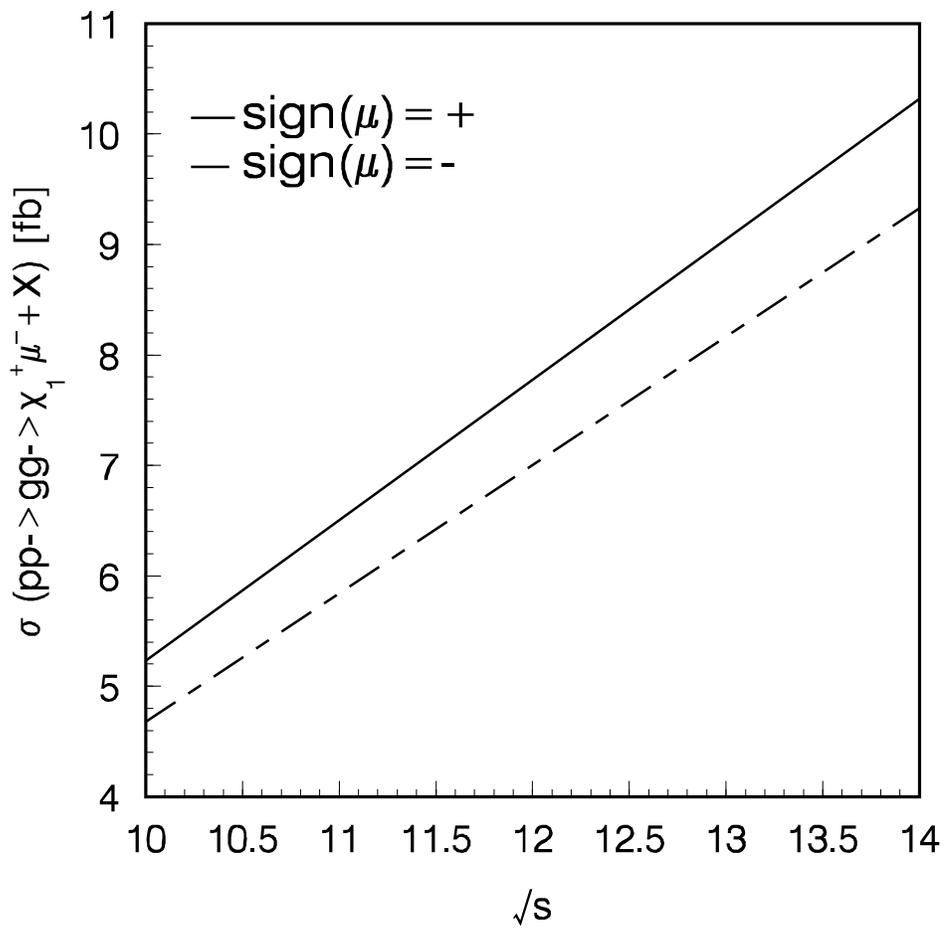
{\bf Fig.4} Total cross section of the process $pp \rightarrow
gg \rightarrow \tilde{\chi}_1^{+}\mu^{-}+X$ as a function of $\sqrt{s}$ at the LHC
in the mSUGRA scenario.

\end{document}